\documentstyle[12pt,aaspp]{article}
\def\beq{\begin{equation}}
\def\eeq{\end{equation}}
\def\ref{\reference}
\def\simge{\mathrel{%
   \rlap{\raise 0.511ex \hbox{$>$}}{\lower 0.511ex \hbox{$\sim$}}}}
\def\simle{\mathrel{
   \rlap{\raise 0.511ex \hbox{$<$}}{\lower 0.511ex \hbox{$\sim$}}}}
\input{psfig}
\eqsecnum
\begin{document}
\title{Hard X-rays from Emission Line Galaxies and the X-ray 
Background: A Test for Advection Dominated Accretion with Radio Observations}

\author{Insu Yi$^1$ and Stephen P. Boughn$^{1,2}$}
\affil{$^1$Institute for Advanced Study, Olden Lane, Princeton, 
NJ 08540; yi@sns.ias.edu}
\affil{$^2$Department of Astronomy, Haverford College, Haverford,
PA, 19041; sboughn@haverford.edu}

\begin{abstract}

Recent studies of the cosmic X-ray background (XRB) have suggested the 
possible existence of a population of relatively faint sources with hard 
X-ray spectra; however, the emission mechanism remains unclear. If the hard 
X-ray emission is from the radiatively inefficient, advection-dominated 
accretion flows (ADAFs) around massive black holes in galactic nuclei, 
X-ray luminosity and radio luminosity satisfy the approximate relation
$L_R\sim 7\times 10^{35}(\nu/15GHz)^{7/5}(M/10^7M_{\odot})(L_x/10^{40} 
erg~s^{-1})^{1/10} ~erg~s^{-1}$ where $L_R=\nu L_\nu$ is the radio 
luminosity at frequency $\nu$, $M$ is the mass of the accreting black hole,
and $10^{40} \simle L_x\simle 10^{42} erg~s^{-1}$ is the 2-10 keV X-ray 
luminosity. These sources are characterized by inverted radio spectra 
$I_\nu \propto \nu^{2/5}$.  For example, an ADAF X-ray source with luminosity
$L_x\sim 10^{41} erg~s^{-1}$ has a nuclear radio luminosity of
$\sim 4\times 10^{36}(M/3\times 10^7M_{\odot}) ~erg~s^{-1}$ at $\sim 20$ GHz 
and if at a distance of $\sim 10(M/3\times 10^7M_{\odot})^{1/2}~Mpc$ 
would be detected as a $\sim 1mJy$ point radio source. High frequency 
($\sim 20 GHz$), high angular resolution radio observations provide an
important test of the ADAF emission mechanism. 
Since $L_R$ depends strongly on black hole mass and only 
weakly on X-ray luminosity, the successful measurement of nuclear radio 
emission could provide an estimate of black hole mass. Because the X-ray 
spectra produced by ADAFs are relatively hard, sources of this emission are 
natural candidates for contributing to the hard, $>2$ keV, background.

\end{abstract}


\section{Introduction}

Recent progress in the study of the cosmic X-ray background (XRB) has 
shown that discrete sources probably account for the entire XRB
(e.g. Hasinger et al. 1993). Although broad line QSOs account for 
$\sim 30\% -50\%$ of the 0.5-2.0 keV XRB (Boyle et al. 1994, Geogantopoulos 
et al. 1996), at higher energies ($>$ 2 keV), the QSO spectrum with energy
spectral index $\alpha_x \sim 1.2$ is too soft to account for the XRB for which
$\alpha_x \sim 0.4$ (Gendreau et al. 1995). Moreover, QSOs are 
significantly more clustered than the diffuse component of the XRB 
(Geograntopoulos et al. 1993, Soltan \& Hasinger 1994). The emission from 
Seyfert nuclei is somewhat harder, $\alpha_x \sim 0.7$, but still too soft to 
account for the XRB unless there is a large population of highly absorbed 
Seyferts ($N_H > 10^{22} cm^{-3}$). Comastri et al. (1995) have 
successfully modeled the amplitude and general shape of the XRB spectrum with 
a unified active galactic nuclei (AGN) model that includes such sources; 
however, the characteristic 
reflection feature in the obscured Seyfert spectrum (e.g. Ghisellini et al. 
1994) has not been observed in the XRB (Gendreau et al. 1995). X-ray spectra 
of starburst galaxies are too soft and their luminosities too low to make 
a significant contribution (Della-Ceca et al. 1996). 

One possible resolution to this problem would be a 
population of ``X-ray galaxies'' which are relatively faint but have
intrinsically hard spectra. This possibility has become especially 
interesting since the recent discovery of narrow emission 
line galaxies with nuclear X-ray luminosities $\sim 10^{41-43} erg~s^{-1}$ 
(Georgantopoulos et al. 1996, Griffiths et al. 1996). Such sources are
$\sim 100$ times more luminous than ``normal'' field galaxies (Fabbiano 1989) 
and yet fainter than typical AGN. If a significant fraction 
of narrow emission line galaxies are similar X-ray sources then they could 
make a substantial contribution to the XRB.  The emission mechanism in these
sources is still unknown.

If nuclear emission in X-ray galaxies were to arise from accretion disks around
massive black holes, i.e. the standard AGN mechanism, and if the central black
hole masses are similar to those of AGN, then the mass accretion rates
would be orders of magnitude lower than for typical AGN (e.g. 
Frank et al. 1992). On the other hand, the relatively low X-ray luminosity could
be due to the low radiative efficiency expected in hot advection-dominated 
accretion flows (ADAFs) 
(e.g. Rees et al. 1982, Narayan \& Yi 1995 and references therein). 
In the latter case, the accretion rates could be 
similar to those of AGN (e.g. Yi 1996). This possibility is especially 
interesting since X-ray emission spectra from ADAFs are intrinsically hard
enough to account for the XRB (Di Matteo \& Fabian 1997). We suggest that 
these two possible X-ray emission mechanisms can be distinguished by the
radio emission which is inherent to ADAFs. The radio detection of such sources
at distances $\simle 10Mpc$ would have important implications on the 
masses of black holes as well as on the XRB. 

We adopt the the following 
dimensionless variables; mass of the black hole $m=M/M_{\odot}$, 
cylindrical radius from the black hole $r=R/R_s$ 
($R_s=2GM/c^2=2.95\times 10^5m~cm$ is the Schwartzschild radius),
and mass accretion rate ${\dot m}={\dot M}/{\dot M}_{Edd}$ where
${\dot M}_{Edd}=L_{Edd}/\eta_{eff} c^2=1.39\times 10^{18} m~ g/s$ is the 
Eddington accretion rate assuming a 10\% efficiency, $\eta_{eff}=0.1$. 

\section{Galaxies as a Source of the X-ray Background}

The cosmic X-ray background intensity at energy $E$ is given by
\beq
{dI(E)\over dE}={c\over 4\pi H_0}\int_0^{z_c}dz{j[E(1+z),z]\over
(1+z)^2(1+\Omega_0z)^{1/2}}
\eeq
where $\Omega_0$ is the usual cosmological density parameter,
$j(E,z)$ is the comoving volume emissivity, 
and $H_0=100h~ km~ s^{-1}~ Mpc^{-1}$ is the Hubble constant. 
For discrete sources with individual differential
luminosity $dL/dE$ and comoving number density $n$, 
$j[E(1+z),z]=n(z)dL/dE(z)|_{E(1+z)}$. We assume that the sources
exist up to redshift $z=z_c$.
If the emissivity evolution is driven by the pure luminosity evolution 
(i.e. without significant spectral or number density evolution), 
one can roughly estimate the number density required to account for 
the observed X-ray background. 
For pure luminosity evolution of the form $L\propto (1+z)^K$ and a 
power-law spectrum $dL/dE\propto E^{-\alpha_x}$,
the integrated X-ray background between the energies $E_1$ and 
$E_2$ is
\beq
I_{XRB}=\int_{E_1}^{E_2} dE {dI\over dE}={cnL_{12}\over 2\pi H_0
(2K-2\alpha_x-3)}\left[(1+z_c)^{K-\alpha_x-3/2}-1\right]
\eeq
where
$L_{12}=\int_{E_1}^{E_2}dE(dL/dE)$ and $\Omega_0=1$ has been assumed for 
simplicity. Therefore,
\beq
nL_{12}={2\pi H_0I_{XRB}(2K-2\alpha_x-3)\over c}
\left[(1+z_c)^{K-\alpha_x-3/2}-1\right]^{-1}.
\eeq
The observed X-ray background is fitted by (e.g. Gruber 1992)
\beq
{dI\over dE}=7.877\left(E\over 1keV\right)^{-0.29}\exp(-E/41.13keV)
~~keV~s^{-1}~cm^{-2}~sr^{-1}~keV^{-1};
\eeq
and, therefore, $I_{XRB}\approx 30 keV~s^{-1}~cm^{-2}~sr^{-1}$ for $E_1=2keV$
and $E_2=10keV$. If $K=3$ and $z_c=2$ (e.g. Almaini et al. 1997), 
with $\alpha_x=0.4$ and $h=0.65$, then $nL_{12}\sim 6 \times10^{38} 
erg~s^{-1}~Mpc^{-3}$, 
which is consistent with the local X-ray emissivity measured 
by Miyaji et al. (1994).  To account for $50\% $ of the $2-10~keV$ XRB in
terms of low luminosity, $L_x \sim 10^{41} erg~s^{-1}$, sources requires
a comoving number density of $n\sim 3 \times 10^{-3} Mpc^{-3}$ which is
comparable to the density of $L^{*}$ galaxies (e.g. Peebles 1993).

There is already evidence that X-ray emission from faint galaxies
contributes significantly to the XRB. Almaini et al. (1997) cross-correlated 
unidentified X-ray sources from 3 deep ROSAT fields and optically faint 
galaxies and concluded that galaxies
with B-band magnitudes less than 23 account for $\sim 20$\% of all X-ray 
sources to a flux limit of $4\times 10^{-15} erg~s^{-1}~cm^{-2}$ 
in the 0.5-2 keV range.
They also found that the X-ray luminosity of these faint galaxies evolves as 
$\propto (1+z)^{3.22\pm 0.98}$ with a comoving emissivity at the present 
epoch of $j_0\approx (3.02\pm 0.23)\times 10^{38} h~ erg~s^{-1}~Mpc^{-3}$ 
in the ROSAT band. 
If integrated to $z_c=2$, $80\pm 20$\% of 
the 0.5-2keV XRB and $40\pm 10$\% of the total XRB are accounted for.
This tentative conclusion emphasizes the need for better understanding of
the X-ray emission mechanism from faint sources.

\section{Radio/X-ray Luminosity Relation}

\subsection{ADAFs in Galactic Nuclei}

When the density of the accretion flow is sufficiently low,
the radiative cooling rate becomes smaller than the viscous heating rate.
As a result, the dissipated accretion energy is not efficiently radiated away 
but kept as internal heat and advected inward with the accreted plasma
(Rees et al. 1982, Narayan \& Yi 1995 and references therein). 
In advection-dominated accretion flows (ADAFs), ions reach a very high 
temperature (close to the virial temperature), the energy transfer from ions 
to electrons becomes slow, and the radiative cooling by electrons remains very 
inefficient. Suppose ions are heated by viscous dissipation at a rate $q_{+}$ 
per unit volume and that viscous heating is characterized by the usual 
viscosity parameter $\alpha$. We assume  $\alpha = 0.3$ (e.g. Frank et al. 1992,
Narayan \& Yi 1995); although, the following results are not very sensitive 
to $\alpha$ unless $\alpha\ll 0.1$. The ADAF emission depends on the ratio 
${\dot M}/\alpha$ so solutions with different $\alpha$'s are equivalent if
${\dot M}$ is appropriately rescaled.
The viscous heating rate is
balanced by the energy advection rate $q_{adv}$ per unit volume and the rate 
of energy transfer to electrons $q_{e+}$ per unit volume; 
\beq
q_{+}=q_{adv}+q_{e+}.
\eeq
When the radiative cooling is inefficient, the accreted plasma becomes
hot and the internal (thermal and magnetic) pressure becomes comparable to
the kinetic pressure.  The dependence of emission on the magnetic field
is more important (e.g. Rees 1982, Narayan \& Yi 1995).
We adopt the equipartition condition that the magnetic 
pressure is equal to $1/2$ the total internal pressure.
Then the  magnetic field in the ADAF is (Narayan \& Yi 1995)
\beq
B\approx 1.1\times 10^4 m_7^{-1/2}{\dot m}_{-3}^{1/2} r^{-5/4}~~ Gauss
\eeq
where $m_7=m/10^7$ and ${\dot m}_{-3}={\dot m}/10^{-3}$.
Hot electrons in magnetic fields radiate through the emission of synchrotron 
photons, the Comptonization of synchrotron photons, 
and bremsstrahlung for which we take $q_{sync}$, $q_{C}$, and
$q_{br}$ as the corresponding rates per unit volume. The synchrotron
emission gives rise to radio and sub-millimeter emission, the
Comptonization of synchrotron photons produces optical and X-ray photons,
and bremsstrahlung contributes only to the X-ray luminosity.
Balancing the heating rate $q_{e+}$ and the total cooling rate, i.e.,
\beq 
q_{e+}=q_{sync}+q_C+q_{br},
\eeq
and simultaneously solving eq. (3-1), we self-consistently determine 
the electron temperature of the ADAF and radiative luminosities in the radio 
and X-ray (for details, see Narayan \& Yi 1995). 
ADAFs exist for mass accretion rates ${\dot m}<{\dot m}_{crit}
\approx 0.3\alpha^2\sim 10^{-1.6}$ (Rees et al. 1982, Narayan \& Yi
1995). The bolometric luminosity of the ADAF, $L_{ADAF}$, is roughly given
by $L_{ADAF}\sim \eta L_{Edd}$ where $\eta\sim 10{\dot m}^2$ 
(Narayan and Yi 1995).
We assume that an ADAF extends from the inner radius of $r_{min}=3$ and
take the outer extent of the flow as $r_{max}\gg r_{min}$.
In this regime, the results depend very weakly on $r_{max}$. 

\subsection{Radio/X-ray Luminosity Relation from ADAFs}

Figure 1 shows radio luminosity 
$\nu L_{\nu}$ at four different frequencies (1.4, 5, 15, 20 GHz) as a function 
of X-ray luminosity in two X-ray bands (0.5-2, 2-10 keV). 
For each black hole mass, luminosities are calculated for ${\dot m}
=10^{-4}-10^{-1.6}$. The lower value of ${\dot m}$
roughly corresponds to the limit above which the electron heating
is dominated by the Coulomb coupling (e.g. Mahadevan 1996). The upper limit in 
principle corresponds to ${\dot m}_{crit}$ for $\alpha=0.3$. 
We note, however, that the exact physical nature of the ADAF for
${\dot m}\sim {\dot m}_{crit}$ is poorly understood. Therefore, luminosity
curves for ${\dot m}\sim {\dot m}_{crit}$ (the far right hand part of the curves in
Figure 1; see also Figure 2) are suspect (Narayan \& Yi 1995).
In each panel, from bottom to top, the black hole mass increases as
$3\times 10^6, 10^7, 3\times 10^7, 10^8, 3\times 10^8 M_{\odot}$.
The curves are steeper at low luminosities and flatter at high luminosities.
This trend is caused by the change in the dominant X-ray emission mechanism.
At lower luminosities, the X-ray luminosity is mainly contributed
by bremsstrahlung; whereas, at high luminosities Compton scattering 
becomes important (e.g. Rees et al. 1982, Narayan \& Yi 1995).

The results shown in Figure 1 can be qualitatively understood as follows.
Optically thin synchrotron emission extends to the self-absorption
frequency, $\nu_s$, at which the Rayleigh-Jeans blackbody emission equals the 
optically thin synchrotron emission.  Therefore, the radio luminosity at 
frequency $\nu$ comes from a radial region where $\nu \approx \nu_s(R)$, 
i.e. $L_{\nu}(\nu) \approx (8\pi^2 k/c^2) R^2T_e\nu^2$ where $T_e$ is
the electron temperature and
\beq
\nu_{s}(r)\approx 9\times 10^{11} m_{7}^{-1/2}{\dot m}_{-3}^{1/2}
T_{e9}^2 r^{-5/4}~~ Hz
\eeq
(Narayan \& Yi 1995, Mahadevan 1996).  In terms of the ADAF parameters
(Rees et al. 1982, Narayan \& Yi 1995, Mahadevan 1996)
\beq
L_R(\nu) \equiv \nu L_{\nu}^{sync}\approx 2\times 10^{32} x_{M3}^{8/5} m_7^{6/5}
{\dot m}_{-3}^{4/5} T_{e9}^{21/5} \nu_{10}^{7/5}~~ erg~s^{-1}
\eeq
where $\nu_{10}=\nu/10^{10} Hz$ and $x_{M3}=x_M/10^3\sim 1$ is the dimensionless
synchrotron self-absorption frequency (Narayan \& Yi 1995). 
Both $T_e$ and $x_{M3}$ vary slowly with $R$ in the inner regions of the ADAF
(Narayan \& Yi 1995, Narayan et al. 1995). The highest radio emission frequency
$\nu_{max}$ arises from the innermost radius of the accretion flow
$r_{min}\sim 3$, i.e.,
\beq
\nu_{max}\approx 2\times 10^{11} m_7^{-1/2} {\dot m}_{-3}^{1/2} T_{e9}^2~~ Hz
\eeq
and the lowest radio emission frequency is expected from the outermost radius of
ADAF $r_{max}\gg r_{min}$,
\beq
\nu_{min}=\nu_{max}(r_{max}/r_{min})^{-5/4}.
\eeq

When the accretion rate ${\dot m}$ is substantially lower than 
${\dot m}_{crit}$, the X-ray emission comes mainly from the bremsstrahlung 
emission (Narayan \& Yi 1995).  Compton
scattering contributes significantly when the mass accretion rate
approaches the critical rate ${\dot m}_{crit}\sim$ a few $\times 10^{-2}$. 
The X-ray luminosity due to the bremsstrahlung emission is
\beq
L_x(\nu) \equiv \nu L_{\nu}^{brem}
\approx 2\times 10^{39} m_7 {\dot m}_{-3}^2 T_{e9}^{-3/2}\nu_{18}
\exp[-4\times 10^{-2} \nu_{18}/T_{e9}]~~ erg~s^{-1}
\eeq
where $\nu_{18}=\nu/10^{18}Hz$ and we have assumed that $kT_e$ is smaller than 
the electron rest energy.  
The Compton luminosity results from the upscattering of the synchrotron
photons and can be approximated by
\beq
\nu L_{\nu}^{Compt}\approx \nu_{max}^{\alpha_c}L_{\nu}^{sync}
(\nu_{max})\nu^{1-\alpha_c}~~ erg~s^{-1}
\eeq
where the index of the Compton scattering spectrum 
$\alpha_c=-\ln\tau_{es}/\ln A$, the electron scattering depth 
$\tau_{es}\approx 5\times 10^{-2}{\dot m}_{-3}$, 
and the scattering amplification factor $A\approx 1+0.7T_{e9}$
(Narayan \& Yi 1995, Mahadevan 1996, and references therein).
Compton scattering is an important mechanism for X-ray emission for
${\dot m}\simge 10^{-3}$ while bremsstrahlung dominates for 
${\dot m}\simle 10^{-3}$.

In the bremsstrahlung dominated regime, 
$L_x\propto m {\dot m}^2$ and $L_R\propto m^{8/5} {\dot m}^{6/5}$. 
These simple scalings are valid when the X-ray photon energy is sufficiently 
smaller than $kT_e$ (cf. eq. 3-8). $T_e$ is approximately independent of
${\dot m}$ and $m$, and $x_M\propto ({\dot m}m)^{1/4}$ (Mahadevan 1996). 
Then, since ${\dot m}\propto (L_x/m)^{1/2}$, we immediately arrive at 
$L_R\propto m L_x^{3/5}$ which qualitatively accounts for the low X-ray
luminosity, steeper parts of the curves in Figure 1.
On the other hand, when 
${\dot m}\simge 10^{-3}$, the X-ray photons in the 0.5-10 keV range
are contributed primarily by Compton up-scattered synchrotron
photons. The luminosity of synchrotron photons up-scattered $N$ times into
the X-ray range is roughly $\propto L(\nu=\nu_{max}) (A \tau_{es})^{N}\propto 
{\dot m}^{N+1}$. For Compton-scattered photons at energies $\sim 2-10 keV$, 
$N>2$. Therefore, $L_x \propto {\dot m}^{N+1}$ and the dependence on 
${\dot m}$ is stronger than in the bremsstrahlung dominated regime.
Thus ${\dot m}\propto L_x^{1/(N+1)}$ and $L_R\propto L_x^{6/5(N+1)}$ 
which accounts for the flatter parts of the curves in Figure 1.
While this argument describes the qualitative shape of the luminosity relations,
more detailed calculations give an even smaller slope at high $L_x$.

The typical radio spectrum peaks at frequencies $>20 GHz$ (eqs. 3-5,3-6) so 
the radio to X-ray luminosity ratio increases with radio frequency.
The radio/X-ray luminosity relation shown in Figure 1 can be approximated by
\beq
L_R=\nu L_{\nu} \approx 7\times 10^{35} m_7 L_{x,40}^{1/10}
(\nu/15GHz)^{7/5}~~ erg~s^{-1}
\eeq
where $L_{x,40}=L_x/10^{40} erg~s^{-1}$ is the X-ray luminosity in the 2-10 keV
band and we have approximated the exponents to the closest simple fractions. 
Despite the change in slope of the luminosity curves in Figure 1, 
this simple radio/X-ray luminosity relation is valid to within a factor of 2
in the luminosity range $10^{40} erg~s^{-1}\simle L_x\simle 10^{42} erg~s^{-1}$.
Since $L_R$ is strongly dependent on $m$ and only mildly dependent on 
$L_x$, the observed radio luminosity provides an imporant constraint on the 
mass of the central black hole.  On the other hand, the inverted
radio spectrum provides an important signature of the ADAF mechanism and
indicates the importance of high frequency ($\nu \simge 5~GHz$) observations.

\subsection{Accretion Flows for High Mass Accretion Rates}

When the accretion rate becomes higher than ${\dot m}_{crit}$, 
the low radiative efficiency ADAFs
no longer exist (Narayan \& Yi 1995). These flows differ from ADAFs 
mainly due to their high radiative efficiency. 
If the accretion flows of the low luminosity galactic nuclei are similar to
those of high luminosity Seyferts and QSOs (e.g. Frank et al. 1992), 
the high efficiency implies a very low mass accretion rate. 
For instance, $m\sim 10^7$ with $L_x\sim 10^{41} erg~s^{-1}$
requires ${\dot m}\sim 10^{-4}$ whereas the ADAFs
would require ${\dot m}$ roughly two orders of magnitude higher.
Since the high efficiency accretion flows are most likely to occur 
in the form of the thin disk (e.g. Frank et al. 1992), a clear correlation
between the radio emission and the X-ray emission comparable to
eq. (3-10) is not expected.

\section{Nearby Candidates for ADAF X-ray Sources}

Radio emission provides a characteristic signature for ADAF sources; however,
the fluxes are relatively low and undetectable for distant sources. 
For example,
an ADAF source with X-ray luminosity $L_x \sim 10^{41}erg~s^{-1}$ and mass
$M \sim 3 \times 10^7 M_{\odot}$ at $z=1$ has a $20GHz$
flux of $\sim 10^{-8}~Jy$.  In addition, resolving small nuclear 
regions at this distance is problematic.  On the other hand, if ADAF sources 
are responsible for a significant fraction of the XRB then their required 
number density, 
$\sim 3\times 10^{-3}~Mpc^{-3}$ (see \S2), is a significant fraction of the
density of field galaxies and they should be easily observable in the 
local universe. If an X-ray bright galaxy contains an ADAF with
$L_x\sim 10^{41} ~erg~s^{-1}$, the 20 GHz radio luminosity is expected
to be $\sim 4\times 10^{36} (M/3\times 10^7M_{\odot}) erg~s^{-1}$. 
Such a galaxy would be detected as a $\simge 1mJy$ radio source out to 
a distance of $\sim 10(M/3\times 10^7M_{\odot})^{1/2} Mpc$.
High resolution 10 and 20 GHz observations of nearby sources 
would test the prediction that the radio spectra are inverted and careful
X-ray observations (e.g. by ASCA) would test the predicted
hardness of the X-ray spectrum.  In fact, a large effort to characterize the 
spectrum of LINERs and low-luminosity AGNs at many wavelengths is already 
underway (Ho 1997) and should provide a test of the general 
characteristics of the ADAF mechanism in these sources.
Recently Falcke et al. (1997) reported the result of their 15 GHz radio 
survey of 48 nearby LINERs optically observed by Ho et al. (1995).
Nuclear radio emission above the $1~mJy$ flux limit was detected in 25\% of 
the LINERs. The observed luminosities are the right order of magnitude but 
further observations at different radio frequencies would be very useful. 

\subsection{Galactic Center Source Sgr A$^*$} 
Sgr A$^*$ in our Galactic center has an X-ray luminosity 
$<5\times 10^{35} erg~s^{-1}$ and perhaps $L_x\sim 10^{35} erg/s$
(Koyama et al. 1996) and a 20 GHz radio luminosity of $\sim 10^{34} erg~s^{-1}$. 
If the black hole 
mass is $\sim 3\times 10^6M_{\odot}$ (Eckart \& Genzel 1996), then the
observed luminosity ratio is within a factor of $\simle 2$ of that predicted for
an ADAF source. Sgr A$^*$ and several other sources are plotted in Figure 2
along with theoretical curves of the luminosity ratio, $L_R /L_x$.  
We conclude that galactic nuclei similar 
to ours could indeed harbor an ADAF (Narayan et al. 1995).  On the other hand, 
extragalactic nuclei with such low luminosities would be extremely 
difficult to detect even if they are relatively nearby.

\subsection{NGC 4258}
For a galactic nucleus with a black hole mass $m\simge 10^7$, 
which is often adopted for Seyferts, the ADAF mechanism implies a significantly
higher radio luminosity than that of Sgr A$^*$.  
Also plotted in Figure 2 is NGC 4258 which has an X-ray luminosity of 
$\sim 4\times 10^{40} erg~s^{-1}$  (Makishima et al. 1994) and a black 
hole mass $\sim 4\times 10^7M_{\odot}$ (Miyoshi et al. 1995).  From 
equation (3-10), the expected $15~GHz$ radio luminosity is 
$\sim 3\times 10^{36} erg~s^{-1}$. Turner \&  Ho (1994) report a marginally 
resolved, compact nuclear radio source at $15~GHz$ in NGC 4258.
Assuming a distance of $\sim 7.5Mpc$ (Sandage 1996), this implies a point source
luminosity of $\simle 3\times 10^{36} erg~s^{-1}$.   We conclude that it
is possible that NGC 4258 is powered by an ADAF (Lasota et al. 1996).

\subsection{LINERs}
NGC 4258 belongs to a group of the emission line galaxies, LINERs
(e.g. Osterbrock 1989). Ho et al. (1995) find that nearly one-third of all 
nearby galaxies with B magnitude $\le 12.5$ are LINERs. Moreover, a 
substantial fraction of them contain broad line regions which suggest that 
these sources may be powered by accretion flows similar to those in AGN 
(Filippenko \& Sargent 1985).  Due to relatively low X-ray luminosities and
hard X-ray emission, LINERs are prime candidates for ADAFs.  
It is, however, unclear whether the accretion
occurs in the form of an ADAF or a high efficiency accretion disk
as in Seyferts (e.g. Lasota et al. 1996). Since the LINERs are classified 
solely based on the optical line ratios, 
the LINERs as a class may be highly heterogeneous. It cannot
be ruled out that some of the galaxies have ADAFs and some have
low luminosity versions of the AGN accretion flows. It is tempting to
speculate that the Seyfert type LINERs could have higher X-ray luminosities
but relatively lower nuclear radio luminosities than ADAF type LINERs.
In this regard, the ADAF radio/X-ray luminosity relation is
particularly promising in distinguishing the X-ray emission mechanism in
emission line galaxies.  For Seyferts with thin disc accretion and radio
jets, the extended radio emission could be much larger than indicated by
equation (3-10).

From a cursory search of the literature we identified 8 LINERs
(NGC 3031, 3079, 3627, 3628, 4258, 4486, 4736, 5194) 
with both measured X-ray fluxes and high angular resolution ($\sim 1''$), 
high frequency ($\simge 10GHz$) radio fluxes.  
The ratios of radio luminosity to X-ray luminosity for these sources are 
plotted in Figure 2 along with ADAF curves for black hole masses ranging
from $10^6$ to $10^9~M_{\odot}$. For two sources (NGC 3627 and 4736)
we computed the $2-10~keV$ fluxes from the $0.5-2~keV$ ROSAT fluxes assuming
an $\alpha_x = 0.6$ power law.  The error in this extrapolation
factor is $< 2$ for a range of spectra from hard to soft.
We also computed the $15~GHz$ radio flux of NGC 3627 from the
reported $8.1~GHz$ flux and of NGC 4486 from the reported $22~GHz$ flux 
using the predicted $\nu ^{2/5}$ power law (equation 3-5). 
The distances to these galaxies were taken from Freeman 
et al. (1994); Sandage (1996); Lehnert \& Heckman (1995); Sandage 
\& Tammann (1987); van den Bergh (1996); Baan \& Haschick (1983)
and also have significant uncertainties. 
Also in Figure 2, we show the luminosity
ratios of 5 typical radio-quiet Seyfert galaxies for comparison.

Even $\sim 1''$ resolution is far too large to resolve the radio emitting
region of ADAFs.  For example, at 10 Mpc $1''$ corresponds to 50 pc compared
to the expected $<1~pc$ size of the ADAF emitting region.  To the extent that
there are other radio sources, e.g. HII regions or radio jets, 
in the radio beam, then the 
ratios of Figure 2 should be considered upper limits.  Of course the resolution
of the X-ray observations is even worse; however, it seems unlikely that there
are other X-ray sources with luminosities $> 10^{40}~erg~s^{-1}$ within $50~pc$
of the nucleus.  If there are such sources then they would have the effect of
artificially lowering the luminosity ratio.  Knowledege of the X-ray and radio
spectra would help clarify these issues.
 
We caution that the list of sources is neither unbiased nor complete. 
We do not intend the plot to be
strong evidence for the existence of the ADAF mechanism in these sources;
rather, only to be suggestive that this might be the case.  In any case,
the uncertainty of black hole masses in these systems coupled with the
strong dependence of radio luminosity on black hole mass preclude
a detailed comparison of observations and predictions. Nevertheless,
most of the sources in Figure 2 appear to be consistent with 
the predicted ADAF luminosity ratios for black hole masses $m\sim 10^7-10^8$
(NGC 3079 and NGC 4486 are separately discussed below).
It is gratifying that the black hole masses are close to those often
invoked for Seyfert-type AGN (e.g. Padovani 1989). 

\subsection{Seyfert Galaxies}
Low luminosity Seyferts with $L_x\simle 10^{43} erg~s^{-1}$ could also be
powered by ADAFs. The five examples of Seyferts shown in Figure 2 are accounted
for by the ADAF emission if they contain black holes with masses
$\sim 10^8-10^9M_{\odot}$. 
If their radio emission is from an ADAF, the positions
of the Seyferts would imply that the sources are accreting at higher
accretion rates closer to the theoretical maximum ${\dot m}_{crit}$
than the LINERs shown. On the other hand, if the Seyferts 
contain high radiative efficiency flows, the mass accretion rates and black 
hole masses could be substantially lower than those estimated in Figure 2.
Then the radio emission is most likely due to a radio jet-like emission
mechanism. In this case, high resolution radio observations could reveal the
radio jets. NGC 1068 and 4151 do appear to contain
jet-like features (e.g. Wilson \& Ulvestad 1982). NGC 1068 has a relatively
strong compact nuclear radio source at the subarcsecond scale (Gallimore et al.
1996), which results in a much higher radio/X-ray flux ratio than the rest of
the Seyferts shown in Figure 2. The radio emission from an ADAF can explain
such a high ratio only if $M\sim 10^9M_{\odot}$. If NGC 1068's black
hole mass is typical of Seyferts, i.e. $<10^8 M_{\odot}$, the relatively
high radio luminosity might indicate small-scale radio jet activity.
However, given the small linear scale $\simle 5pc$ (Gallimore et al.
1996), radio emission from an ADAF seems more likely. If the latter is
the case, then the black hole mass is estimated to be close to
$\sim 10^9M_{\odot}$.  It should be noted that starburst activity in 
NGC 1068 contributes to the nuclear, hard X-ray
emission (e.g. Ueno et al. 1994).  NGC 4151 is a variable X-ray source so we
used the mean $2-10$ keV flux (Papadakis \& McHardy 1995).

\subsection{Radio Bright Sources: Need for High Resolution Radio Observations}
The high radio to X-ray ratio of NGC 3079 also implies 
a black hole mass as high as $\sim 10^9M_{\odot}$. 
This is also the case for the giant 
elliptical galaxy NGC 4486 (M87).  For M87 the dynamical estimates
of the black hole mass, $\sim 3\times 10^9M_{\odot}$, 
(Harms et al. 1994) are indeed consistent with the position of the source in 
Figure 2.  NGC 4486 does have a radio jet; however, the radio
flux quoted here is from a VLBI measurement (Moellenbrock et al. 1996) and 
corresponds to a spatial resolution of $< 1~pc$. The similarity
between the type 2 Seyfert NGC 1068 and the two LINERs NGC 3079, 4486
is intriguing and it may suggest a common emission mechanism in the
two different types of active galaxies with comparable luminosities.
 
Generally, radio-loud galactic nuclei populate the upper right
region of Figure 2.  NGC 3079, 4486, and 1068 are relatively radio loud and 
so it is crucial to extract the compact nuclear radio emission
(i.e. excluding contribution from extended radio lobes).
High angular resolution radio observations at two frequencies would
improve the situation considerably by replacing the upper limits with
actual detection of point sources and by testing whether or not the sources
have the predicted inverted spectrum. 
Measurements of the X-ray spectral index would further
test the ADAF emission mechanism.

\section{Summary and Discussion}

Recent developments in the study of the X-ray background indicate the
necessity of relatively faint, hard-spectrum X-ray sources. These sources
could be either heavily obscured low luminosity AGN or X-ray bright
emission line galaxies.  The implied luminosity of these X-ray bright 
galaxies is similar to that of LINERs and other narrow emission 
line galaxies and it has been suggested that nearly 1/3 of all 
nearby galaxies are LINERs (Ho et al. 1995).  A prime candidate for the 
emission mechanism of X-ray bright galaxies is
ADAF onto a central, massive black hole.  While classical, 
bright AGN are expected to have weak radio emission (unless they have
radio jets), ADAF sources exhibit high frequency radio emission from a hot 
synchrotron-emitting plasma.  
The radio luminosities of ADAFs are modest (observations at 20 GHz would
only be able to detect them to distances of $\simle 30~Mpc$); however, the
characteristic inverted radio spectrum provides an observational test 
for the ADAF paradigm.  If ADAFs are unambiguously detected, the 
radio/X-ray luminosity relation provides a direct estimate of the 
central black hole mass.  Several nearby LINERs have radio to X-ray 
luminosity ratios that are consistent with the ADAF predictions; 
however, more observations are needed to test the predictions 
presented in this paper.

An important consequence of the inherently inefficient ADAF mechanism is
the implied growth rate for the masses of the central black holes.
If a typical source $2-10~keV$ luminosity at $z=0$ is 
$L_x\sim 10^{41} erg~s^{-1}$,
and if the sources undergo the luminosity evolution of the form
$(1+z)^3$ up to redshift 2 (as suggested by Almaini et al. 1997), 
then at $z=2$, the source has a luminosity of 
$\sim 3\times 10^{42} erg~s^{-1}$ which is similar to the luminosities of some
X-ray bright galaxies (Griffiths et al. 1996 and
references therein). The approximate minimum black hole mass for these sources
consistent with ADAF emission can be estimated by equating this
luminosity with the maximum bremsstrahlung luminosity
expected from the ADAF accretion flow.  The maximum $2-10~keV$ band 
limited flux is $\sim 10^{42} m_7 ~erg~s^{-1}$ (Narayan \& Yi 1995). Then 
the the minimum black hole mass at $z=2$ is $\sim 3\times 10^7M_{\odot}$. 
An estimate of the current black hole mass for these sources at $z=0$ 
is obtained by assuming that these sources continue to accrete at a rate 
${\dot m}\sim {\dot m}_{crit}\sim 2\times 10^{-2}$ from $z=2$ to $z=0$
(cf. Yi 1996). The resulting black 
hole mass, $\sim 5\times 10^7M_{\odot}$, is similar to the 
typical black hole mass assumed for Seyfert galaxies and the black hole 
mass observed in NGC 4258. This rough estimate is not intended to be
a prediction of the current mass spectrum of galactic black holes but 
rather to check whether or not the ADAF mechanism violates current estimates
of central black hole masses.

We have hypothesized that ADAFs
around central black holes in galaxies generates a substantial component of the
hard X-ray background.  This requires that a significant fraction of galaxies
harbor central black holes with ADAF flows.  However, even if
the number density of ADAF sources is too small to contribute significantly to
the XRB (cf. Di Matteo \& Fabian 1997), 
ADAFs are still interesting for X-ray bright emission line galaxies and
the observational tests suggested here are useful probes of the X-ray emission 
mechanism.

\acknowledgments
We would like to thank Ramesh Narayan for information on the spectral states of 
accreting black holes and Roeland van der Marel for helpful discussions on black
hole masses. This work was supported in part by the SUAM and Monell Foundations
and NASA grant\# NAG 5-3015.

\clearpage

\noindent

\vfill\eject
\clearpage

\centerline{Figure Captions}

\vskip 0.3cm

Figure 1: X-ray luminosity vs. radio luminosity for four black hole masses in
erg/s units. In each panel, the black hole mass increases as $3\times 10^6$,
$10^7$, $3\times 10^7$, $10^8$, and $3\times 10^8$ solar masses from bottom to
top curves. In the top panels, the X-ray luminosity is for the
2-10 keV band and in the bottom panels, the X-ray luminosity refers to
the 0.5-2 keV range. From left to right, the radio frequency increase
from 1.4 GHz to 20 GHz.

\vskip 0.3cm
 
Figure 2: The ratios of 15 GHz luminosity to 2-10 keV luminosity for 8 LINERs
(NGC 3031, 3079, 3627, 3628, 4258, 4486, 4736, 5194), the Galactic Center
radio source Sgr A$^*$, and 5 Seyfert galaxies 
(NGC 1068, 3227, 4151, 5548, 4388). Due to finite angular
resolution in the radio the ratios should be considered as upper limits.  
The theoretical curves are (from bottom to top) for black hole masses
$\log M/M_{\odot}$ of $6.0$ to $9.0$ as marked with the numbers.
The X-ray fluxes are from Serlemitsos, Ptak, \& Yaqoob (1997); 
Brinkmann, Siebert, \& Boller (1994); Makishima et al. (1994); 
Fabbiano, Kim, \& Trinchieri (1992); Reynolds et al. (1996);
Smith \& Done (1996); Papadakis \& McHardy (1995).  
The radio fluxes are from Turner \& Ho (1994); Carral, Turner, \& Ho (1990); 
Saikia, et al. (1994); Spencer \& Junor (1986); Moellenbrock et al. (1996);
Gallimore et al. (1996). The dotted lines indicate the luminosity ratios 
for given ${\dot m}=10^{-4}, 10^{-2}, 2.5\times 10^{-2}$.


\begin{references}
\reference Almaini, O. et al. 1997, MNRAS, submitted
\reference Bann, W. A. \& Haschick, A. D. 1983, AJ, 88, 1088
\reference Boyle, B. J. et al. 1994, MNRAS, 271, 639
\reference Boyle, B. J. et al. 1995, MNRAS, 272, 462
\reference Brinkmann, W., Siebert, J., \& Boller, Th. 1994, A\&A,281,355
\reference Carral, P., Turner, J. L., \& Ho, P. T. P. 1990, ApJ, 362, 434
\reference Comastri, A. et al. 1992, ApJ, 384, 62
\reference Della-Ceca, R. et al. 1996, ApJ, 469, 662
\reference Di Matteo, T. \& Fabian, A. C. 1997, MNRAS, 286, 393
\reference Eckart, A. \& Genzel, R. 1996, Nature, 383, 415
\reference Fabbiano, G. 1989, ARA\&A, 27, 87
\reference Fabbiano, G., Kim, D. W., \& Trinchieri, G. 1992, ApJS, 80, 531 
\reference Falcke, H. et al. 1997, preprint
\reference Filippenko, A. V. \& Sargent, W. A. 1985, ApJS, 57, 503
\reference Frank, J., King, A. R., and Raine, D. 1985, Accretion 
Power in Astrophysics, Cambridge: Cambridge University Press
\reference Freedman, W. L. et al. 1994, ApJ, 427, 628
\reference Gendreau, K. C. et al. 1995, PASJ, 47, L5
\reference Genzel, R., Hollenbach, D., \& Townes, C. H. 1994, Rep. Prog. Phys.,
57, 417
\reference Georgantopoulos, I. et al. 1993, MNRAS, 262, 619 
\reference Georgantopoulos, I. et al. 1996, MNRAS, 280, 276
\reference Gallimore, J. F. et al. 1996, ApJ, 458, 136
\reference Ghisellini, G., Haardt, F., Matt, G. 1994, MNRAS, 267, 743
\reference Griffiths, R. E. et al. 1995, MNRAS, 275, 77
\reference Griffiths, R. E. et al. 1996, MNRAS, 281, 71
\reference Gruber, D. E. 1992, in the Proceedings of "The X-ray Background",
ed. X. Barcons \& A. C. Fabian (Cambridge: Cambridge Univ. Press), p44
\reference Harms, S. N. et al. 1994, ApJ, 435, 35
\reference Hasinger, G. et al. 1993, A\&A, 275, 1
\reference Ho, L. C. 1997, private communication
\reference Ho, L. C., Filippenko, A. V., \& Sargent, W. A. 1995, ApJS, 98, 477
\reference Koyama, K. et al. 1996, PASJ, 48, 249
\reference Lasota, J-P. et al. 1996, ApJ, 462, 142
\reference Lehnert, M. D.\& Heckman, T. M. 1995, ApJS, 97, 89.
\reference Mahadevan, R. 1996, ApJ, 465, 327
\reference Makishima, K. et al. 1994, PASJ, 46, L77
\reference Miyaji, T. et al. 1994, ApJ, 393, 134
\reference Miyoshi, M. et al. 1995, Nature, 373, 127
\reference Moellenbrock, G. A. et al. 1996, AJ, 111, 2175
\reference Narayan, R. \& Yi, I. 1995, ApJ, 444, 231
\reference Narayan, R., Yi, I., \& Mahadevan, R. 1995, Nature, 374, 623
\reference Osterbrock, D. E. 1989, Astrophysics of Gaseous Nebulae and
Active Galactic Nuclei (Mill Valley, CA: University Science Books)
\reference Papakakis, I. E. \& McHardy, I. M. 1995, MNRAS, 273, 923
\reference Peebles, P. J. E. 1993, Principles of Physical Cosmology
(Princeton Univ Press, Princeton, NJ), 122
\reference Padovani, P. 1989, A\&A, 209, 27
\reference Rees, M. J., Begelman, M. C., Blandford, R. D., Phinney, E. S.
1982, Nature, 295, 17
\reference Reynolds, C. S. et al. 1996, MNRAS, 283, 111
\reference Saikia, D. J., Pedlar, S. W., Unger, S. W., \& Axon, D. J. 1994,
MNRAS, 270, 46
\reference Sandage, A. 1996, AJ, 111, 18.
\reference Sandage, A. \& Tammann, G. A. 1997, 
A Revised Shapley-Ames Catalogue of 
Bright Galaxies (2nd ed.), Carnegie Institute of Washington: Washington, DC
\reference Serlemitsos, P., Ptak, A., \& Yaqoob, T. 1997, astro-ph/9701127
\reference Smith, D. A. \& Done, C. 1996, MNRAS, 280, 355
\reference Soltan, A. \& Hasinger, G. 1994, A\&A, 288, 77
\reference Spencer, R. E. \& Junor, W. 1986, Nature, 321, 753
\reference Turner, J. L. \& Ho, P. T. P. 1994, ApJ, 421, 122
\reference Ueno, S. et al. 1994, PASJ, 46, L71
\reference van den Bergh, S. 1996, PASP, 108, 109
\reference Wilson, A. S. \& Ulvestad, J. S. 1982, ApJ, 263, 576
\reference Yi, I. 1996, ApJ, 473, 645
\end{references}
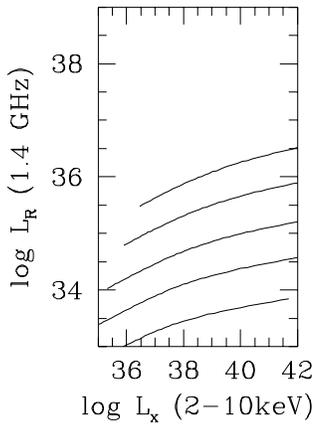
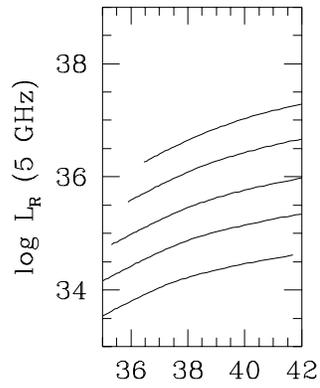
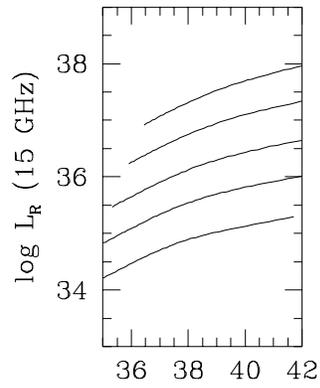
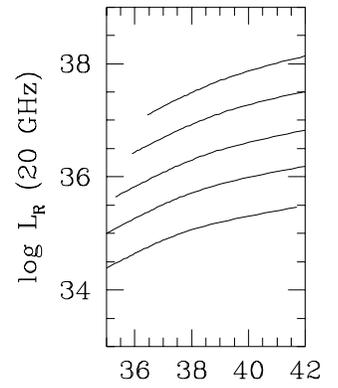
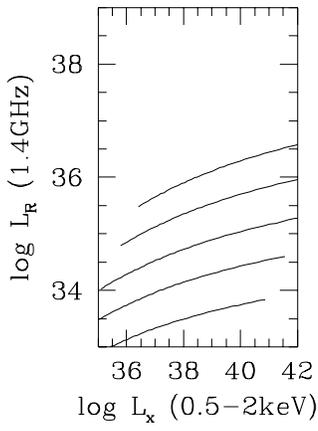
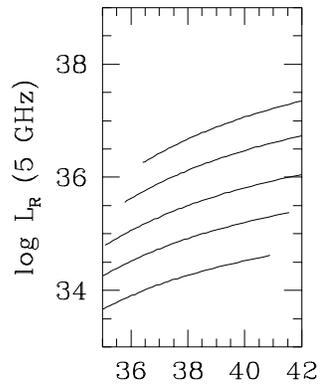
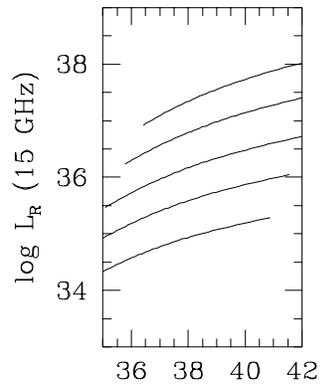
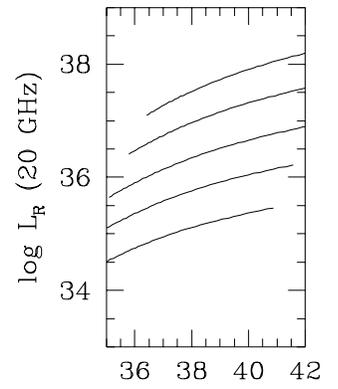
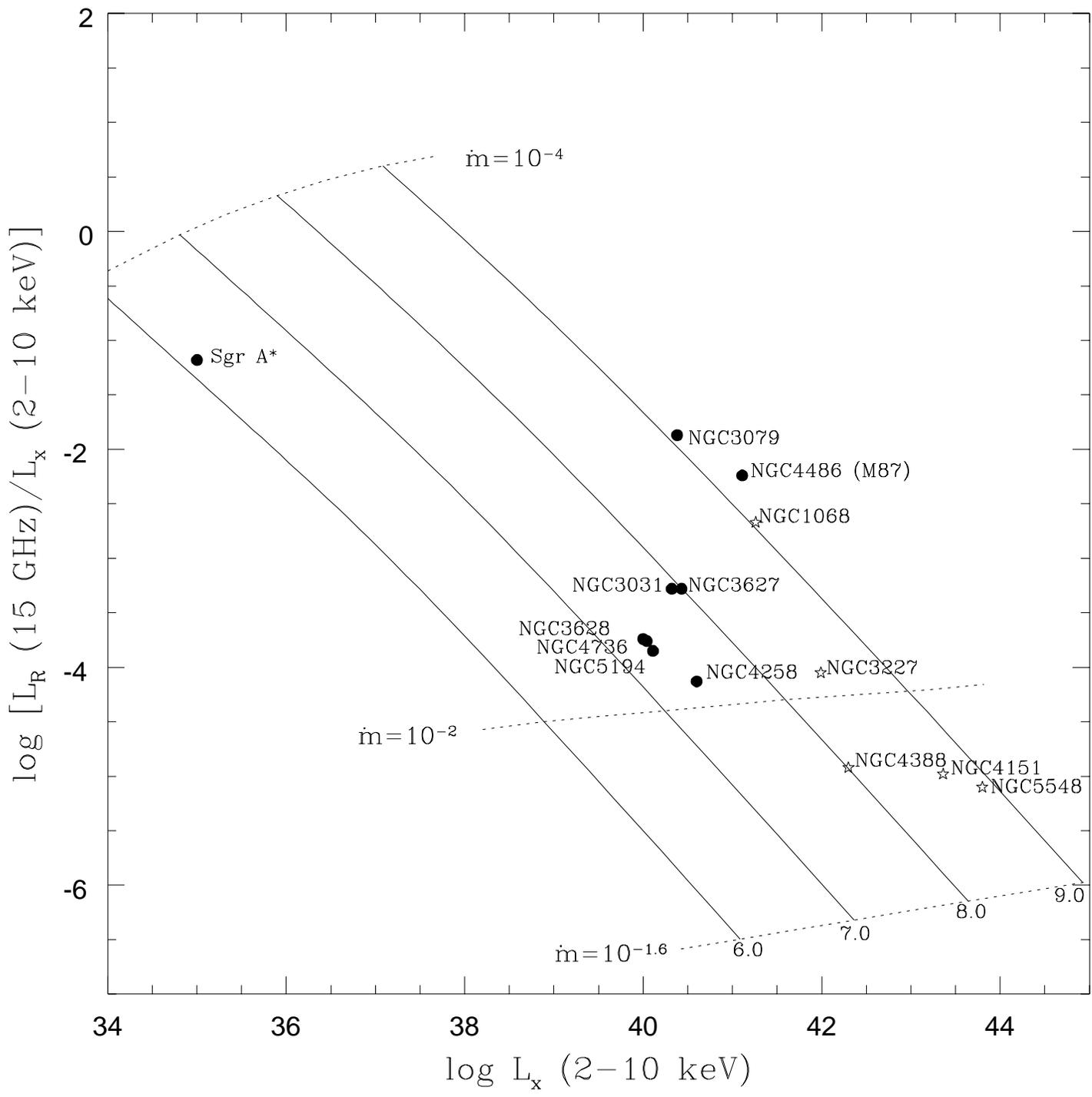
\end{document}